\documentclass[3p,times,procedia]{elsarticle}
\usepackage{nupha_ecrc}

\usepackage[compact]{titlesec}
\titlespacing{\section}{0pt}{1.6ex}{0.8ex}

\volume{00}

\firstpage{1}

\journalname{Nuclear Physics A}

\runauth{}

\jid{nupha}

\jnltitlelogo{Nuclear Physics A}

\usepackage{amssymb}

\usepackage[figuresright]{rotating}
\usepackage{caption}

\newcommand{\s}{\sqrt{s}}
\newcommand{\sNN}{\sqrt{s_{\scriptscriptstyle \rm NN}}}

\newcommand{\pPb}{\mbox{p--Pb}}

\newcommand{\pt}{p_{\rm T}}

\newcommand{\Dz}{{\rm D}^0}
\newcommand{\Ds}{{\rm D}^{*+}}
\newcommand{\Dp}{{\rm D}^+}

\newcommand{\Jpsi}{{\rm J}/\Psi}

\newcommand{\RpPb}{R_{\rm pPb}}

\newcommand{\dphi}{\Delta \mathrm{\varphi}}
\newcommand{\deta}{\Delta \mathrm{\eta}}

\begin{document}

\begin{frontmatter}



\dochead{}

\title{Heavy-flavor correlations and multiplicity dependence in pp and p--Pb collisions with ALICE}


\author{Fabio Colamaria, on behalf of the ALICE Collaboration}

\address{Dipartimento Interateneo di Fisica ‘M. Merlin’ and INFN, Sezione di Bari, Italy}

\begin{abstract}
The production of heavy quarks in pp collisions at the LHC energies provides a reliable test of perturbative QCD calculations. Comparisons of pp and $\pPb$ measurements of their hadronization products allow us to investigate how cold nuclear matter effects affect the heavy-quark production.
We present ALICE measurements of azimuthal correlations of prompt D mesons with charged hadrons in pp collisions at $\s$ = 7 TeV and $\pPb$ collisions at $\sNN$ = 5.02 TeV. We also show the per-event D-meson yields as a function of the charged-particle multiplicity in pp collisions at $\s$ = 7 TeV.
\end{abstract}


\end{frontmatter}


\vspace{0.3cm}
\section{Introduction}
\label{sec:intro}
The study of heavy-flavor production in pp collisions at LHC energies allows us to test perturbative QCD calculations and provides a reference for studies in heavy-ion collisions. Measurements in $\pPb$ collisions help to characterize the effects due to the presence of a nucleus in the collision (cold nuclear matter effects). ALICE provided measurements of $\pt$-differential cross sections for D-meson production at central rapidity in pp~\cite{D_pp7,Ds_pp7,D_pp276} and $\pPb$~\cite{D_pPb} collisions, and of the nuclear modification factor, $\RpPb$. More differential measurements of charm production in pp and $\pPb$ collisions can provide further insight on the above topics.
The study of heavy-flavor production in pp collisions as a function of the charged-particle multiplicity allows us to investigate the interplay between hard and soft QCD processes responsible for particle production in hadronic collisions. It also provides information on the role of multi-parton interactions (MPI) in the heavy-flavor sector.
Moreover, the analysis of angular correlations between heavy-flavor particles and charged particles is a tool to characterize the heavy-quark fragmentation process and is sensitive to their production mechanism. Differences between the measurements in pp and $\pPb$ collisions can give insight on how cold nuclear matter effects affect heavy-quark production and hadronization in $\pPb$ collisions.

\section{D-meson yields as a function of event multiplicity}
\label{sec:multiplicity}

In ALICE (see~\cite{alice} for a complete description of the detector and its performance), D mesons and their charge conjugates are reconstructed from their hadronic decay channels (D$^0\rightarrow$ K$^-\pi^+$, D$^+\rightarrow$ K$^-\pi^+\pi^+$, D$^{*+}\rightarrow$ D$^0 \pi^+\rightarrow$ K$^-\pi^+\pi^+$ and D$_{\rm s}^+\rightarrow \phi\pi^+ \rightarrow$ K$^-$K$^+\pi^+$) and selected on the basis of their displaced decay vertex topology. D-meson daughters are selected exploiting particle identification and track-quality cuts~\cite{D_pp7}.

The self-normalized yields of $\Dz$, $\Dp$ and $\Ds$ are evaluated as a function of the charged-particle multiplicity normalized by its average in multiplicity-integrated events, for different $\pt$ ranges of the D mesons, in pp collisions at $\s$ = 7 TeV. This observable is defined as the ratio of the D-meson per-event yield in a given multiplicity interval over the multiplicity-integrated yield. The number of track segments in the two innermost layers of the Inner Tracking System in $|\eta| < 1$ is used as multiplicity estimator; it is found to be proportional to the density of charged particles at central pseudorapidity.
The fraction $f_{\rm B}$ of D mesons from B-hadron decays (feed-down), evaluated on the basis of FONLL calculations~\cite{FONLL}, is assumed independent of the charged-particle multiplicity, hence it cancels out in the self-normalized yields.
The results obtained for $\Dz$, $\Dp$ and $\Ds$ mesons are in agreement. This allows us to evaluate a weighted average of the measurements for the three mesons, to reduce the statistical uncertainties.

\begin{figure}[!h]
\centering
\begin{minipage}{\linewidth}
  \centering
  $\vcenter{\hbox{\includegraphics[width=.44\linewidth]{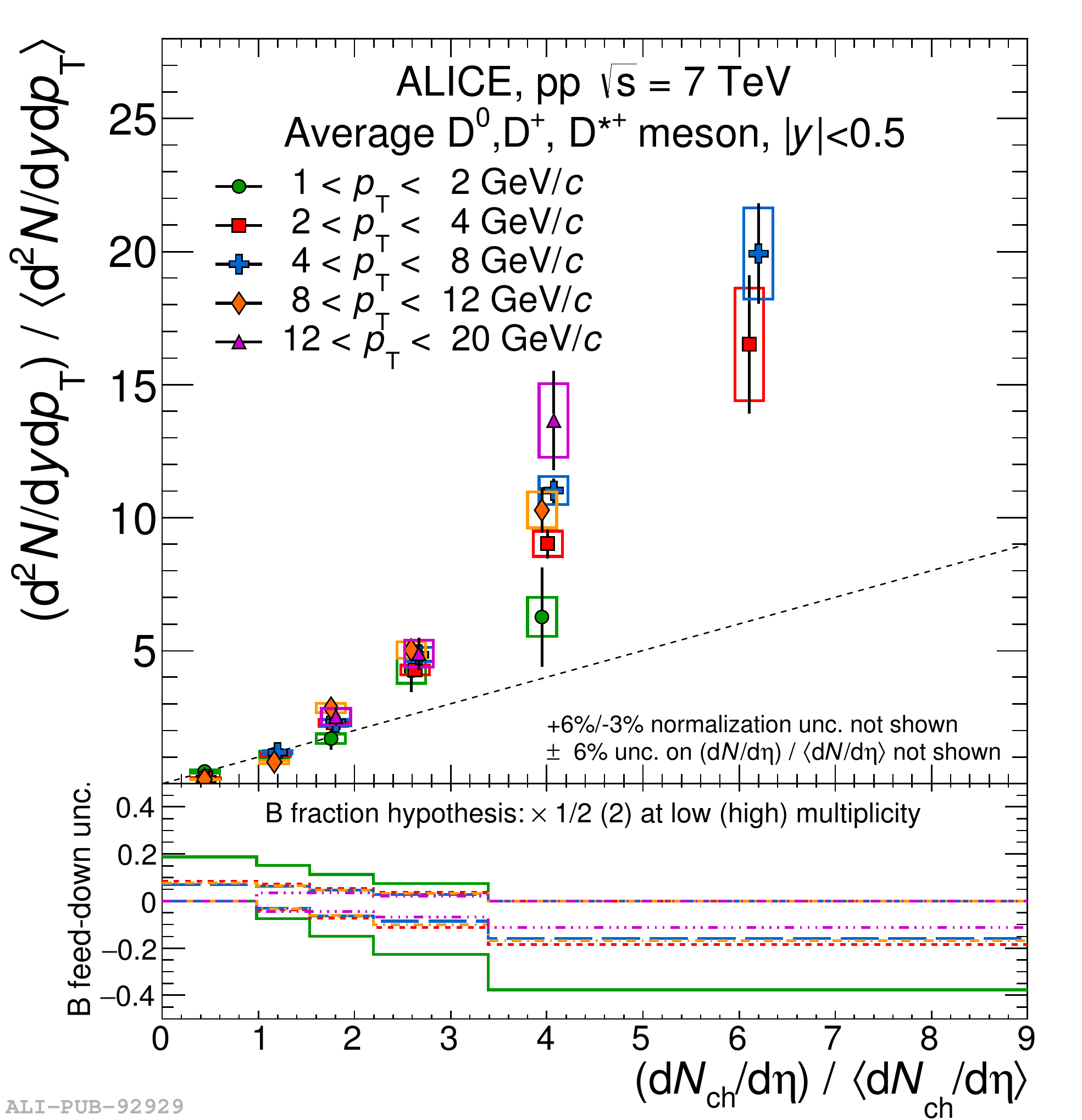}}}$
  $\vcenter{\hbox{\includegraphics[width=.53\linewidth]{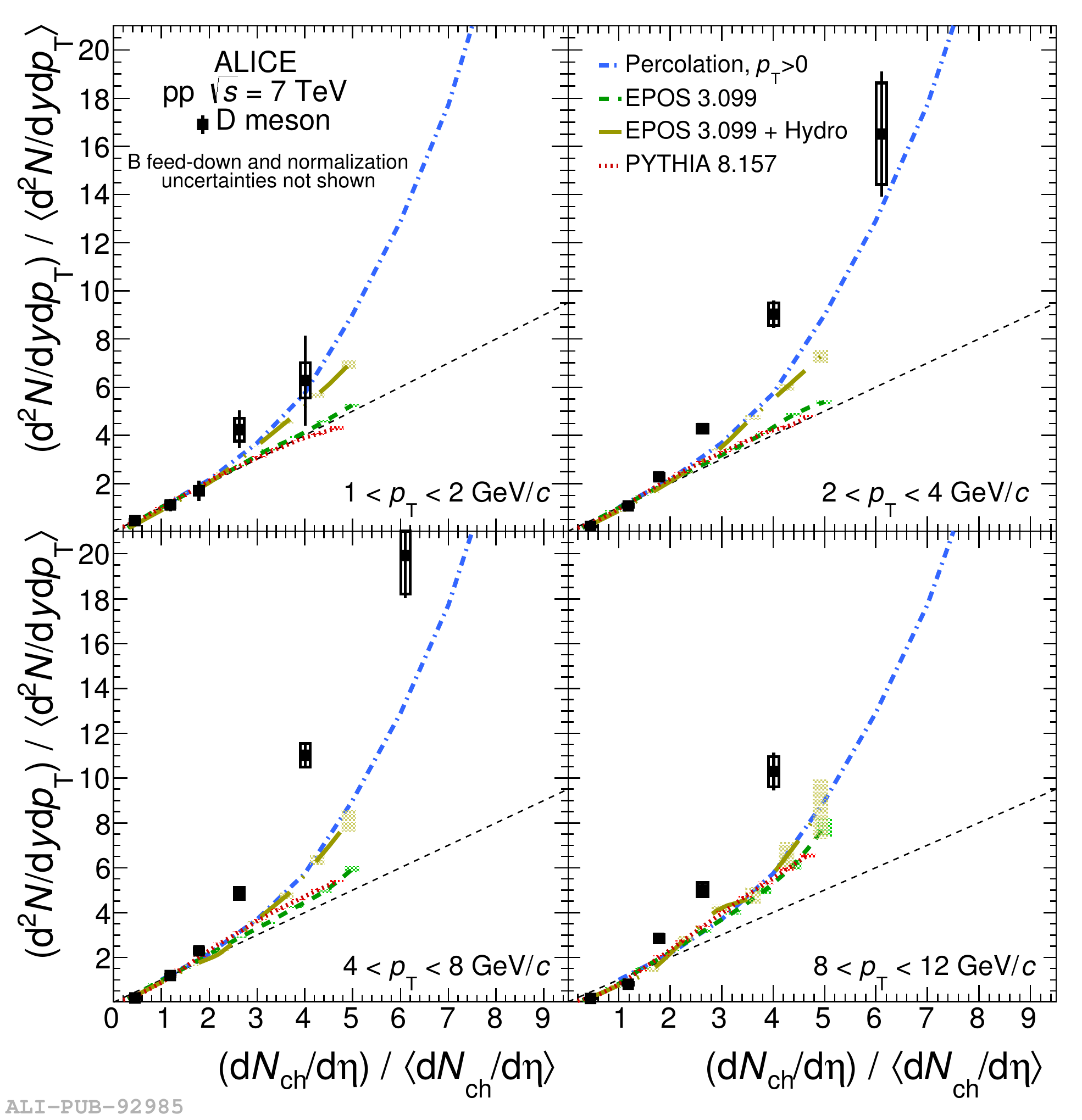}}}$
\end{minipage}
\caption{Left panel: average of $\Dz$, $\Dp$ and $\Ds$ self-normalized yields as a function of the relative charged-particle multiplicity at central rapidity, for five $\pt$ intervals, in pp collisions at $\s$ = 7 TeV. Right panel: comparison of average D-meson self-normalized yields with calculations from models (see text for details).}
\label{fig:1}
\end{figure}

The average D-meson self-normalized yields are shown in the left panel of Fig.~\ref{fig:1}~\cite{DmesonvsMult}, for five D-meson $\pt$ ranges from 1 to 20 GeV/$c$. A faster-than-linear increase of the self-normalized yields with event multiplicity can be observed, with a trend independent of the D-meson $\pt$ within uncertainties. In the right panel of Fig.~\ref{fig:1}, the self-normalized yields are compared to predictions from a percolation model~\cite{perc1,perc2}, EPOS 3~\cite{epos1,epos2} and PYTHIA 8~\cite{pythia8}, all of which include a contribution of MPI to particle production. A qualitative agreement with data is observed for all the models, though at high multiplicities PYTHIA 8 and EPOS 3 without hydrodynamical evolution of the collision seem to underestimate the increasing trend with multiplicity observed in data.

\begin{figure}[!h]
\centering
\begin{minipage}{\linewidth}
  \centering
  $\vcenter{\hbox{\includegraphics[width=.46\linewidth]{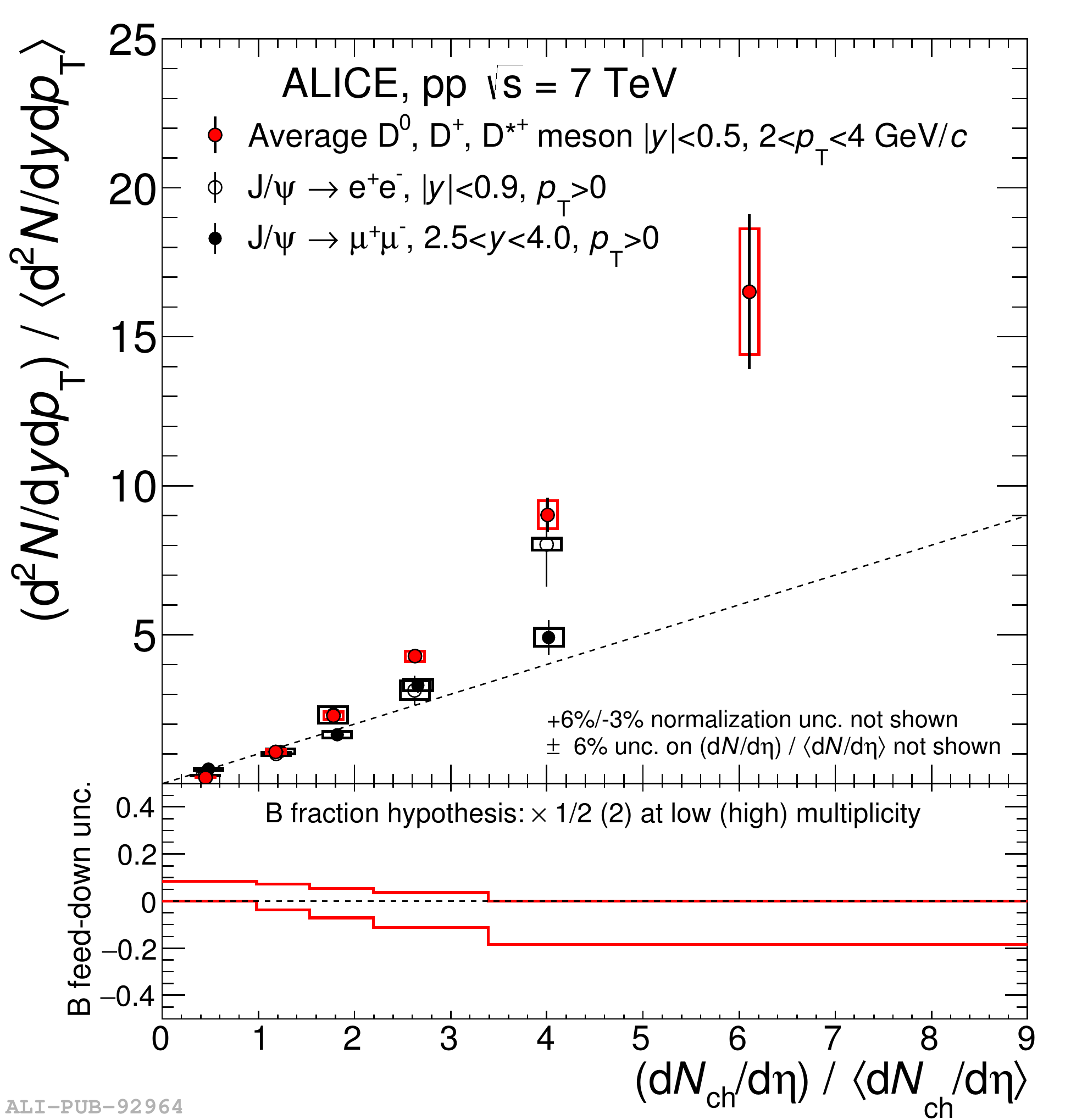}}}$
  $\vcenter{\hbox{\includegraphics[width=.46\linewidth]{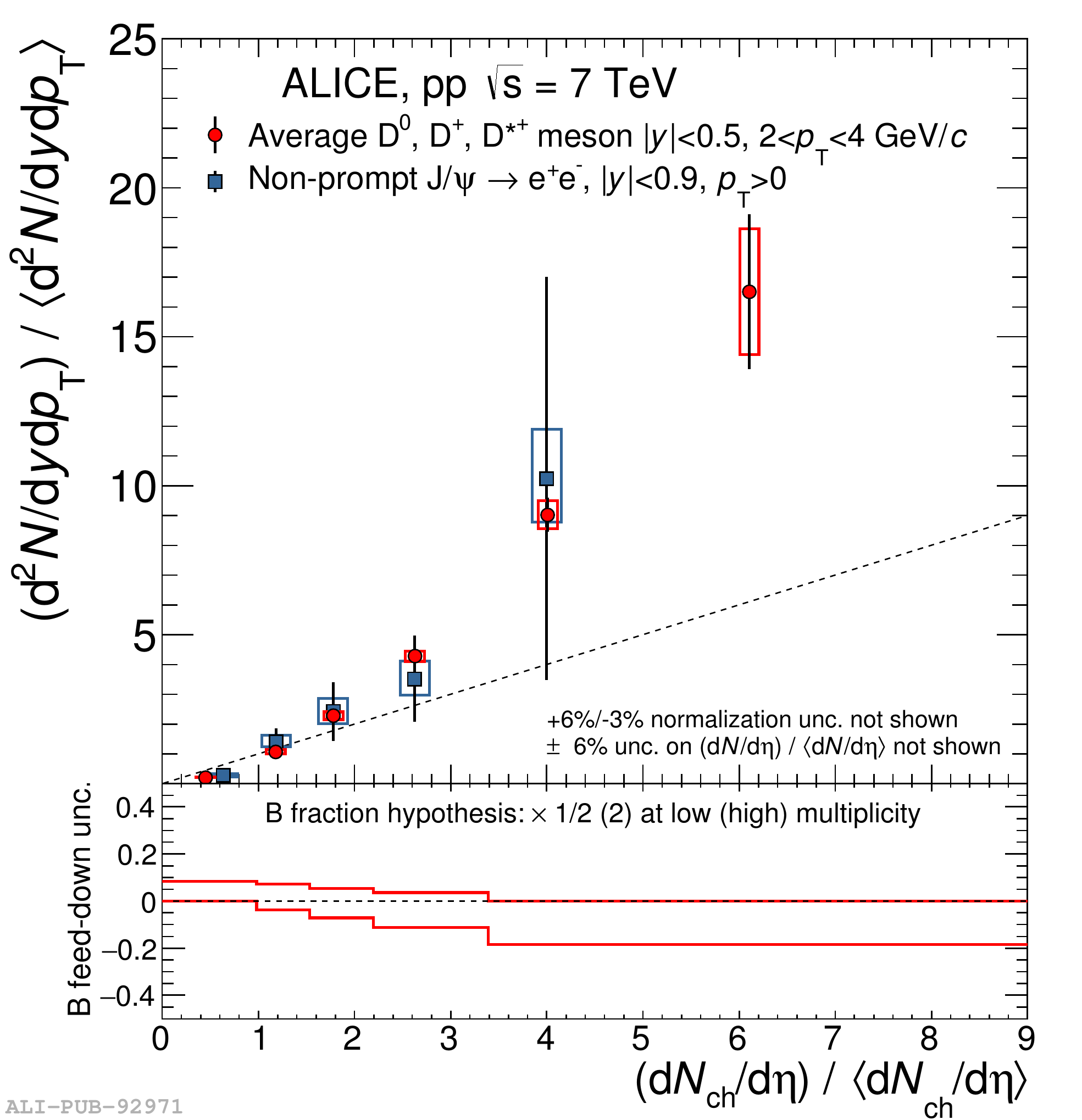}}}$
\end{minipage}
\caption{Comparison of average D-meson self-normalized yields with inclusive $\Jpsi$ at central and forward rapidity (left panel) and with non-prompt $\Jpsi$ at central rapidity.}
\label{fig:2}
\end{figure}

A comparison of the self-normalized yields of D meson and $\Jpsi$ in pp collisions at $\s$ = 7 TeV is shown in the left panel of Fig.~\ref{fig:2}. $\Jpsi$ are reconstructed at central rapidity ($|y| <$ 0.9) in the $e^+e^-$ decay channel and at forward rapidity (2.5 $< y <$ 4) in the $\mu^+\mu^-$ decay channel~\cite{Jpsi}. A similar increase with event multiplicity is found for the open and hidden charm self-normalized yields, suggesting that this behavior is driven by quark production mechanisms, rather than by their hadronization. The same trend is also observed for the production of open charm and open beauty, as shown in the right panel of Fig.~\ref{fig:2}, where the multiplicity dependence of the self-normalized yields of D meson and non-prompt $\Jpsi$ is compared.

\section{D meson-charged particle angular correlations}
\label{sec:correlations}
Two-dimensional angular correlation distributions of $\Dz$, $\Dp$ and $\Ds$ mesons (trigger particles) with charged particles (associated particles) are evaluated in pp collisions at $\s$ = 7 TeV and $\pPb$ collisions at $\sNN$ = 5.02 TeV, for different ranges of the D-meson and associated particle $\pt$.

The contribution due to D-meson combinatorial background is removed by subtracting the correlation distribution evaluated from the sidebands of the D-meson invariant mass distribution. A correction for detector inhomogeneities and limited acceptance is applied exploiting the event-mixing technique. Efficiency corrections for reconstruction and selection of trigger and associated particles are performed.
The contribution of D mesons from B feed-down is subtracted. In particular, the templates of the angular correlations of feed-down D mesons and charged particles are built using PYTHIA simulations~\cite{pythia} and normalized to the amount of feed-down contribution evaluated from FONLL calculations~\cite{FONLL}.
A projection on the $\dphi$ axis, limited to the range $|\deta| <$ 1, is performed. The azimuthal correlation distributions of $\Dz$, $\Dp$ and $\Ds$ mesons are compatible within uncertainties. A weighted average of the three D-meson measurements is hence performed, to reduce the statistical uncertainty on the points.
The per-trigger azimuthal correlation distributions are fitted with two Gaussian functions on top of a constant (baseline), to extract physical observables such as the near-side yield, near-side width and height of the baseline.

\begin{figure}[!h]
\centering
\begin{minipage}{\linewidth}
  \centering
  $\vcenter{\hbox{\includegraphics[width=.48\linewidth]{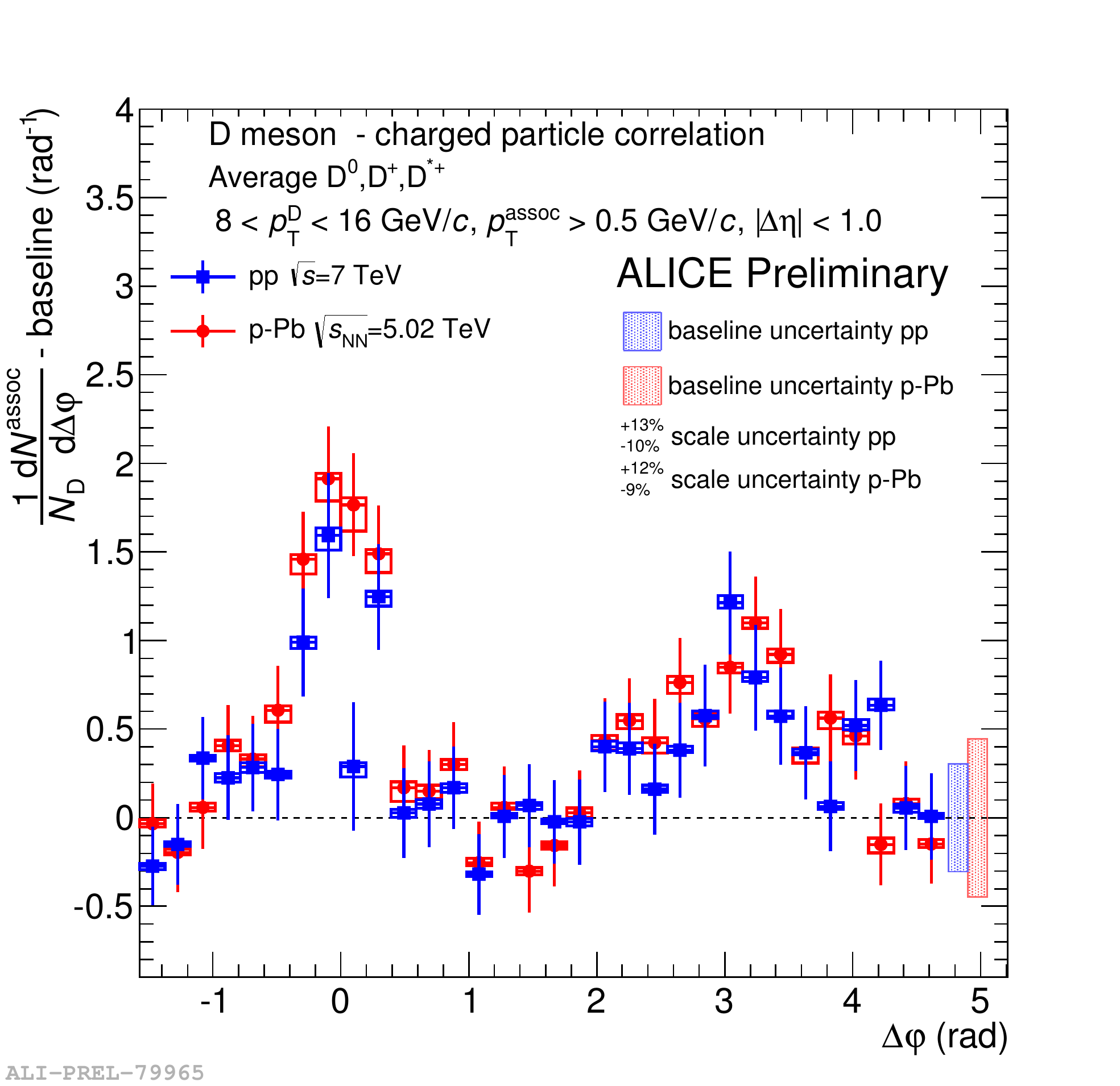}}}$
  $\vcenter{\hbox{\includegraphics[width=.48\linewidth]{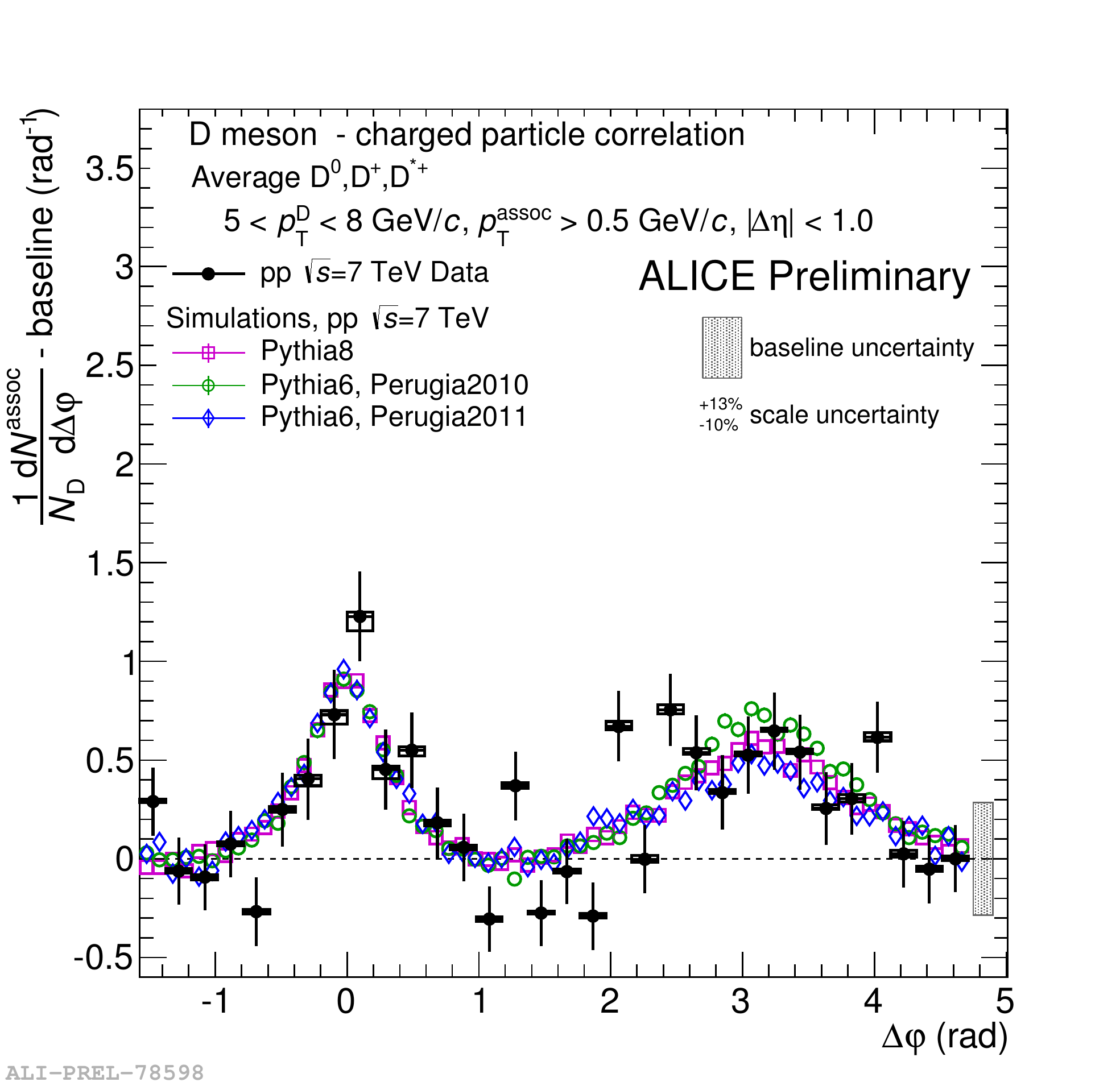}}}$
\end{minipage}
\caption{Left panel: comparison of average D meson-charged particle azimuthal correlation distributions in pp and $\pPb$ collisions, after baseline subtraction, for 8 $< \pt($D$) <$ 16 GeV/$c$ and associated particle $\pt >$ 0.5 GeV/$c$. Right panel: comparison of baseline-subtracted average D meson-charged particle azimuthal correlation distributions in pp collisions and predictions from different tunes of the PYTHIA generator, for 5 $< \pt($D$) <$ 8 GeV/$c$ and associated particle $\pt >$ 0.5 GeV/$c$.}
\label{fig:3}
\end{figure}

\begin{figure}[!h]
  \begin{minipage}[c]{0.47\textwidth}
    \includegraphics[width=\textwidth]{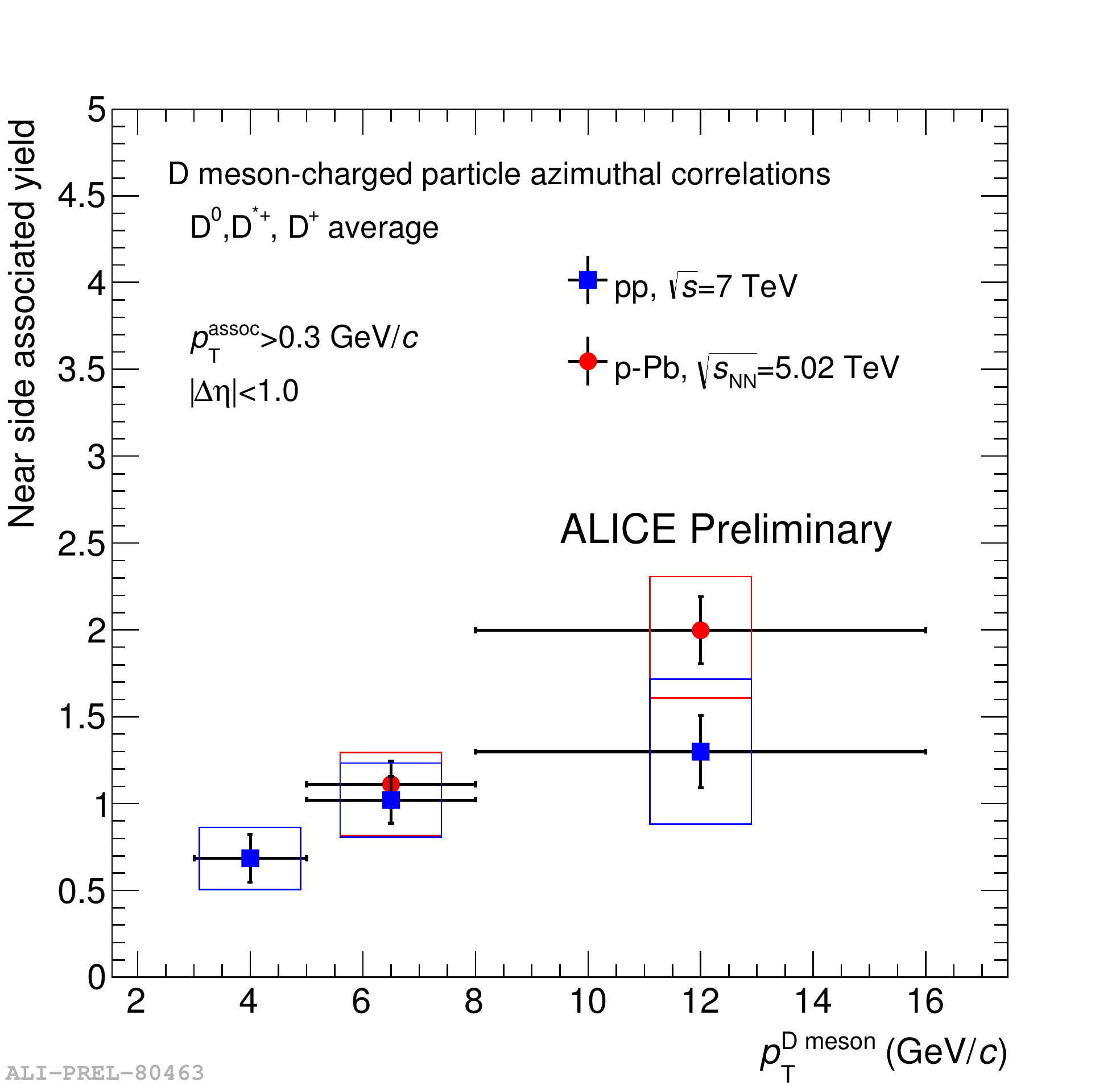}
  \end{minipage}\hfill\hspace{0.5cm}
  \begin{minipage}[c]{0.53\textwidth}
    \caption{Comparison of near-side associated yields extracted in pp and p-Pb collisions as a function of the D-meson $\pt$, for associated particle $\pt >$~0.3 GeV/$c$. The near-side yields are extracted by integrating the near-side Gaussian component of the fit function.} 
    \label{fig:4}
  \end{minipage}
\end{figure}

A comparison of the baseline-subtracted average D meson-charged particle azimuthal correlation distributions, evaluated in pp and $\pPb$ collisions, for 8 $< \pt($D$) <$ 16 GeV/$c$ and associated particle $\pt >$ 0.5 GeV/$c$, is shown in the left panel of Fig.~\ref{fig:3}. Compatibility within the uncertainties of the correlation distributions in the two collision systems is found for all the kinematic ranges. The azimuthal correlation distributions are well reproduced by PYTHIA simulation templates in all kinematic ranges, as shown in the right panel of the same figure for 5 $< \pt($D$) <$ 8 GeV/$c$ and associated particle $\pt >$ 0.5 GeV/$c$, in pp collisions.
A comparison of the trend of the near-side yields as a function of the D-meson $\pt$ in pp and $\pPb$ collisions is shown in Fig.~\ref{fig:4}, for associated particle $\pt >$ 0.3 GeV/$c$. Compatible values of the near-side yields are observed in pp and $\pPb$ collisions. Within the current uncertainties, no modifications of the $\pPb$ yields due to cold nuclear matter effects are observed.

\section{Summary}
The self-normalized D-meson yields as a function of the charged-particle multiplicity, measured by ALICE in pp collisions, show a faster-than-linear increase with multiplicity, in fair agreement with expectations from models including MPI contribution to particle production. The D meson-charged particle azimuthal correlation distributions, measured in pp and $\pPb$ collisions, are in agreement between each other within uncertainties, and are well described by Monte Carlo expectations.





\bibliographystyle{elsarticle-num}
\bibliography{biblio}







\end{document}